\documentclass{article}
 \usepackage{graphics} 
\usepackage{epsfig }
\newcommand{\gevc }{GeV/$c$}

\newcommand{\q }{$q$}
\newcommand{\Deq }{$\Delta q$}
\newcommand{\Deqt }{$\Delta_T q$}

\begin{document}
 \title{COMPASS  Results  on Transverse Single-Spin Asymmetries}
 \author{Anna Martin~\footnote{
Presented at Advanced Studies Institute: Symmetries And Spin (SPIN-Praha-2006)
        19-26 Jul 2006, Prague, Czech Republic,
Proceedings to be published in Czechoslovak Journal of Physics.}\\
 University of Trieste and INFN Trieste \vspace*{0.5cm}\\
 {\em on behalf of the COMPASS Collaboration}}
  
 \maketitle

 \begin{abstract}
New results on single spin asymmetries of charged
hadrons produced in deep-inelastic scattering of muons
on a transversely  polarised $^6$LiD target are presented. 
The data were taken in the years 2002, 2003 and 2004 
with the COMPASS spectrometer
using the muon beam of the CERN SPS at 160~\gevc .
Preliminary results are given for the Sivers asymmetry and for
all the three
``quark polarimeters'' presently used in
COMPASS to measure the transversity distributions.
The Collins and the Sivers asymmetries
for charged hadrons turn out to be
 compatible with zero, 
within the small ($\sim 1$\%) statistical errors, at variance
with the results from HERMES on a transversely 
polarised proton target.
Similar results have been obtained for the two hadron
asymmetries and for the $\Lambda$ polarisation.
First attempts to describe the Collins and the Sivers asymmetries
measured by  COMPASS
and HERMES allow to give a consistent picture of
these transverse spin effects.
 \end{abstract}

{\small
\begin{description}
\item[{\rm {\em PACS:}}] 13.60.-r, 13.88.+e, 14.20.Dh, 14.65.-q
\item[{\rm {\em Key words:}}] transverse single-spin asymmetry, Collins asymmetry, 
Sivers asymmetry, COMPASS
\end{description}
  }

 \section{Introduction} 
The importance of transverse spin effects at high energy is
well known since 30 years, when the first measurements took place.
It is only in the nineties, however, that the experimental and the
theoretical progress allowed to point out their relevance in
the understanding of the nucleon 
structure~\footnote{
For a more detailed description see, f.i.,~\cite{newpap}
and references therein.}.
In that period, several new experiments (HERMES at DESY, COMPASS at CERN,
the RHIC experiments at BNL, ...) were proposed to
 investigate the spin structure of the nucleon
and the transverse spin effects, and important progress was done 
on the theoretical side.
Since that, the activity in the field is continuously growing,
giving a more and more complete picture.

Today, it is well established that 
to fully specify  the quark structure of the nucleon 
at the twist-two level, 
the transverse spin distribution functions \Deqt\  have to be added to the
unpolarised
distributions \q\ and the helicity distributions \Deq .
The ``new'' parton distribution functions (PDF's), also called  ``transversity'',
are related to the
probability that the quark spins are aligned parallel or
antiparallel to the spin of
a transversely polarised nucleon, and are still unknown.
The distributions \Deqt\ are in fact difficult to measure, since 
they are chirally odd and need to be coupled to a chirally odd
partner. In particular, they cannot be measured in
inclusive deep-inelastic scattering (DIS).
They can be measured in\\
- hard scattering of transversely polarised protons on 
transversely polarised protons  looking at Drell-Yan processes
as proposed by the RHIC experiments~\cite{RHIC}. Here, however, the measured
quantity is related to $\Delta_T q \cdot \Delta_T \bar{q}$ which is
expected to be very small;\\
- hard polarised proton antiproton scattering looking again 
at Drell-Yan processes.
This method has been proposed at the new FAIR facility at GSI~\cite{gsi}
and will allow to measure directly $\Delta_T u \cdot \Delta_T u$.
It should be a very clean measurement, but will not allow
to extract informations on the transversity distribution of the other
quarks;\\
- semi-inclusive DIS (SIDIS)
of leptons on transversely polarised nucleons in which 
final state hadrons are  also detected. This method is the
subject of the present talk.

To access the transversity PDF in
SIDIS, one has to measure the quark polarisation, i.e.
to use the so-called ``quark polarimetry''.
Different techniques have been proposed in so far. Three of them
are presently used in COMPASS, namely:\\
- measurement of the single-spin asymmetries (SSA) in the azimuthal 
distribution
of the final state hadrons (the so-called ``Collins asymmetry'');\\
- measurement of the SSA in the azimuthal distribution
of the plane containing the final state  hadron pairs (the so-called 
``two-hadron asymmetry'');\\
- measurement of the polarisation of 
final state $\Lambda$ hyperons (the so-called ``$\Lambda$ polarimetry''),\\
which are described in the following sections.

In the last ten years, the interest for 
the intrinsic transverse momentum of quarks inside a nucleon
or a hadron grew-up considerably, and the transverse momentum dependent
(TMD) distribution functions and fragmentation functions (FF)
are today considered an important ingredient in the structure of
the nucleon.
Some of these TMD distributions can be extracted in SIDIS looking at the
azimuthal distributions of the final state hadrons.
As explained in the next section, this is true
in particular for the so-called ``Sivers asymmetry'', which is, together
with the Collins asymmetry, presently the most studied.
Through the Sivers asymmetry it is possible to access the Sivers PDF,
which takes into account a possible deformation in the distribution
of the quark intrinsic transverse momentum in a transversely
polarised nucleon.
 
 \subsection{Single hadron asymmetries}

 \subsubsection{The Collins asymmetry}
The distributions \Deqt\ may  be
extracted from measurements of the SSA's in the cross-section
of SIDIS of leptons on transversely polarised nucleons.

This method is based on the so-called ``Collins effect'', which 
could be responsible for a left-right asymmetry in the fragmentation
of transversely polarised quarks~\cite{Collins:1993kk}.
For spinless hadrons, the fragmentation function of a polarised quark
is expected to be of the form
$$
D_{T\,q}^{\; \; \; h}(z, \vec{p}_T^{\, h}) = D_q^h(z, {p}_T^{\, h}) + 
    \Delta_T^0 D_q^h(z, {p}_T^{\, h}) \cdot \sin \Phi_C ,
$$
where $\vec{p}_T^{\, h}$ is the  hadron transverse momentum with 
respect to the quark direction, $z$ is the fraction of available energy carried by 
the hadron, $D_q^h(z, {p}_T^{\, h})$ is the usual FF,
and the Collins function $\Delta_T^0 D_q^h$ is the 
spin dependent $T$-odd part of the FF of a transversely 
polarised quark $q$ into a hadron $h$.
The ``Collins angle'' $\Phi_C=\phi_h - \phi_{s'}$ is the difference between
the azimuthal angle of the hadron and the azimuthal angle of the spin
of the fragmenting quark evaluated from the quark momentum.
The Collins function $\Delta_T^0 D_q^h$ has to be measured independently,
and, after some attempts to extract it from the $e^+ e^-$ annihilation data
from the DELPHI experiment at LEP, it is now being measured at BELLE. 
First results~\cite{Abe:2005zx,Ogawa:2006bm} indicate very clearly that the Collins effect is real and
that the spin dependent FF is different from zero.
The Collins effect is also expected to be largest 
for the leading hadron in the current jet, i.e. the hadron with the
highest momentum~\cite{Artru}.

\begin{figure}[tb] %
\begin{center}
\includegraphics[width=0.7\textwidth]{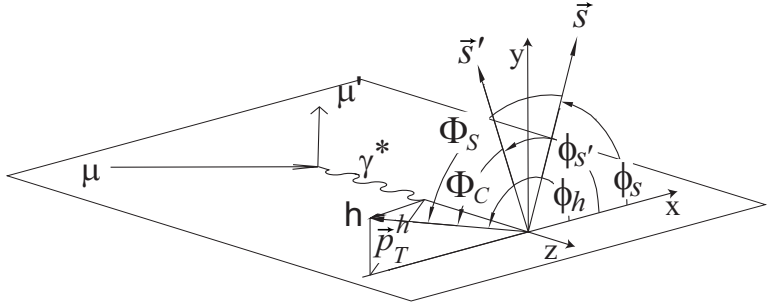}
\caption{Definition of the Collins and Sivers angles.
The vectors $\vec{p}_T^{\, h}, \, \vec{s}$ and $\vec{s}\,'$
are  the hadron transverse momentum and the spins of the initial and struck
quarks respectively.}
\label{fig:angles}
\end{center}
\end{figure}

In SIDIS on a transversely polarised target,
the Collins effect is responsible for a 
measurable asymmetry in the distribution of final state hadrons,
related to a convolution 
of the transversity distributions and the Collins fragmentation functions. 
The
number of hadrons $h$ in a given bin of the Bjorken variable $x$, or of $z$,
or of ${p}_T^{\, h}$ is given by
$$
N^{h \, \pm }(\Phi_C)   =  N^{h \, \pm }_0 \cdot (1 \pm
     A_C^m \cdot \sin \Phi_C)
$$
where $\pm$ refers to the target spin orientation (spin ``up'' and ``down''
in the laboratory system), and
$N^{h \, \pm }_0$ is the mean number of hadrons,
depending on the incident
lepton flux, the acceptance and efficiency of the apparatus and the 
unpolarised cross-section.
The Collins angle $\Phi_C$ is evaluated as if the target spin is always oriented
``up''.
In SIDIS, since $\phi_{s'} = \pi - \phi_s$, where $\phi_s$ is
the azimuthal angle of the initial state quark with respect to the lepton 
scattering plane, it is given by
$$
\Phi_C=\phi_h+\phi_S-\pi \, .
$$
where $\phi_h$ and $\phi_S$ are the azimuthal angles  
with respect to the lepton 
scattering plane of the hadron transverse momentum and  of the initial nucleon
spin respectively,
as illustrated in Fig.~\ref{fig:angles}.

From the angular distribution of the hadrons, one can thus measure 
the asymmetry
$$
    A_C^m  =  P_T \cdot f \cdot D_{NN} \cdot A_{Coll} \, ,
$$
where 
$f$ is the polarised target dilution factor, 
$P_T$ is the target polarisation, and
$D_{NN} = (1-y)/(1-y+y^2/2)$ is the transverse spin
transfer coefficient from the initial to 
the struck quark.
At leading twist in the collinear parton model, the ``Collins
asymmetry'' is given by
$$
   A_{Coll} = \frac {\sum_q e_q^2 \cdot \Delta_T q \cdot 
                  \Delta_T^0 D_q^h}
{\sum_q e_q^2 \cdot q \cdot D_q^h} \, .
$$
Using proton and deuteron (or neutron) targets and selecting different
final state hadrons
and knowing the Collins FF's, it is thus possible separate the contributions 
of the quarks of different flavour and thus extract the 
transversity distributions.

In the most general case in which the quark intrinsic momentum is considered
and
TMD distributions and fragmentation 
functions are used, 
convolution integrals appear in the SIDIS 
cross-section~\cite{aram94,MuTa96}. Still,
assuming Gaussian distributions for the parton and the hadron
transverse momenta in the PDF's and in the FF's, the previous expression for 
the Collins asymmetry is  valid when multiplied by
a factor which depends on mean values of the transverse 
momenta (see f.i.~\cite{Collins:2005ie,efre}).

 \subsubsection{The Sivers asymmetry}
When considering the parton and the hadron
transverse momenta, at leading order, in addition to the transversity
PDF, in the SIDIS cross-section two other PDF's appear in the case
of unpolarised lepton 
beam and transversely polarised target, and one more if the lepton
beam is longitudinally polarised.

The corresponding SSA's depend on different and independent functions of
the azimuthal angles of the target nucleon spin and of the momentum of the
hadron, and can all be
extracted from the same data samples.

Till now only the ``Sivers asymmetry'' has been measured, in addition to the
Collins one.

As suggested by Sivers~\cite{Sivers} there is a possible correlation
between the transverse momentum $\vec{k}_T$ of an unpolarised quark
in a transversely polarised nucleon and the nucleon polarisation vector, 
i.e.  the spin-averaged quark distribution $q(x)$
can be written as
$$
q_{T}(x,\vec{k}_T)= q(x,{k}_T) + | \vec{S}_\perp | \cdot 
            \Delta_0^T q(x,{k}_T)\cdot \sin \, \Phi_{ S}
$$
where the ``Sivers angle''
$\Phi_{ S}= \phi_q -\phi_S$
is the relative azimuthal angle between the quark transverse momentum
$\vec{k}_T$ and the target spin $\vec{S}_\perp$, and
the PDF $\Delta_0^T q(x,{k}_T)$ is the so-called ``Sivers function''.
If the hadron produced in
the fragmentation and the fragmenting quark are collinear, i.e. if
all the hadron transverse momentum originates from the intrinsic 
transverse momentum of the quark in the nucleon, the Sivers angle,
shown in Fig.~\ref{fig:angles}, becomes $\Phi_{ S}= \phi_h -\phi_S$.
The distribution of the number of events in the Sivers angle for the
two target orientations is
then given by
$$
N^{h \, \pm}(\Phi_S)   =  N^{h \, \pm}_0 \cdot (1 \pm
     A_S^m \cdot \sin \Phi_S) \, . \nonumber
$$
In this case the measurable asymmetry is
$$
    A_S^m  =  P_T \cdot f  \cdot A_{Siv} \, .
$$
At leading twist and assuming the final hadron to be
collinear with the fragmenting quark, the ``Sivers asymmetry'' is given by
$$
   A_{Siv} = \frac {\sum_q e_q^2 \cdot \Delta_0^T q \cdot 
                  D_q^h}
{\sum_q e_q^2 \cdot q \cdot D_q^h} \, .
$$
In this case the FF's are known; combining data collected with proton and 
deuteron (or neutron) targets and selecting different
final state hadrons it is thus possible to separate the contributions 
of the quarks of different flavour from SIDIS data only.

As for the Collins asymmetry, 
assuming Gaussian distributions for the parton and the hadron
transverse momenta in the PDF and in the FF, the previous expression for 
the asymmetry is still valid when multiplied by
a factor which depends on the mean values of the transverse 
momenta.

 \subsection{Two-hadron asymmetry}
In accessing transversity via single hadron Collins asymmetry,
one may be sensible to  different effects which might
damp the effect or even spoil the
measurement.
As an alternative method it has been proposed (see~\cite{Artru} and 
references therein)
to measure the ``relative Collins effect'' between two fast particles of the
jet, which allows to measure an asymmetry which is practically
not affected by fluctuations in the fragmenting quark momentum.
Assuming that the $q\bar{q}$ pairs are created in the 
$^3P_0$ state, an asymmetry is expected in a semi-classically model of 
string fragmentation.
In the S-matrix theory, the asymmetry is due to the interference 
between the helicity-flip amplitude and the helicity-conserving one 
(the ``interference fragmentation'')~\cite{Bacchetta:2006un}.

Experimentally, this asymmetry is measured very much like the Collins one,
essentially replacing the azimuthal angle of the hadron $\phi_h$ with the 
azimuthal angle of the plane of the two hadrons $\phi_R$.
In COMPASS $\phi_R$ is defined as the azimuthal angle with respect to
the lepton scattering plane of $\vec{R}_T$, the component transverse to
the virtual photon of the vector
$$
\vec{R}=\frac{z_2 \cdot 
          \vec{p}^{\, h_1} - z_1 \cdot \vec{p}^{\, h_2}}{z_1+z_2} \, ,
$$
where the indexes 1 and 2 refer to the two final state hadrons.

Form the angular distribution in $\phi_{RS} = \phi_R - \phi_h$ it
is possible to measure the asymmetry 
$$
    A_{2h}^m  =  P_T \cdot f \cdot D_{NN} \cdot A_{RS} \, ,
$$
where 
$$
   A_{RS} = \frac {\sum_q e_q^2 \cdot \Delta_T q \cdot 
                  H_q^{2h}}
{\sum_q e_q^2 \cdot q \cdot D_q^{2h}} \, .
$$
Both the spin dependent fragmentation function $H_q^{2h}$, and the 
corresponding spin-averaged FF $D_q^{2h}$, are unknown, and need to be measured
in $e^+e^-$ annihilation or to be evaluated using models.
They are expected to depend on $z=z_1+z_2$ and on the invariant mass
of the two hadrons.

 \subsection{$\Lambda$ polarimetry}
This is one of the first methods proposed to access transversity, and the
favoured one by some authors (see f.i.~\cite{Artru} and~\cite{ans02}).
In this case, the information on the transversity PDF is obtained 
by measuring the polarisation of the current jet $\Lambda$ ($\bar{\Lambda}$)
produced in SIDIS.

The transverse polarisation of the $\Lambda$ measured with respect
to the nucleon spin axis $\vec{S}_{\perp}$ is related to the transversity 
PDF by
\begin{eqnarray}
   P_{\Lambda}^S = f\cdot P_T \cdot D_{NN} \cdot
               \frac {\sum_q e_q^2 \cdot \Delta_T q \cdot 
                  \Delta_T D_q^{\Lambda}}
{\sum_q e_q^2 \cdot q \cdot D_q^{\Lambda}} \, ,
\end{eqnarray}
 where $\Delta_T D_q^{\Lambda}$ is the transversely polarised fragmentation
function of the quark $q$ into a $\Lambda$. It  needs to be measured 
independently, and  is interesting by itself.

Introducing  in the $\Lambda$ rest frame 
the angle $\theta _S^*$ between  $\vec{S}_{\perp}$ 
and the momentum of the proton emitted in the $\Lambda$ decay,
$P_{\Lambda}^S$ can be measured by the distribution of the 
number of events, since
$$
   N^{\pm}(\theta _S^*) \simeq 1 \pm \alpha 
        \cdot  P_{\Lambda}^S \cdot \cos \theta _S^* \, ,
$$
where $\alpha = 0.642 \pm 0.013$ is the decay
asymmetry parameter.
The sign $\pm$ refer to the two possible orientations
of the target spin ($\vec{S}_{\perp}$ is defined pointing upward
in the laboratory system for both target spin orientations).

The only disadvantage of this measurement, which makes it less popular
than the previous ones, is that, as expected, its statistical efficiency 
as a quark polarimeter
is poor, due to the relatively small abundance of  
weak-decaying hyperons in quark jets.

 \section{The COMPASS experiment}
COMPASS is a fixed target experiment 
which was proposed to CERN in 1996 to investigate 
hadron structure and hadron spectroscopy using both hadron 
and muon high-energy beams from the SPS.
Apart from a short pilot run in 2004 with a pion beam, in so far
the experiment  has focussed on the study of the
spin structure of the nucleon, taking data from 2002 to 2004 
with a $\mu^+$ beam 
and a polarised deuteron target.

The experiment has been run at a muon energy of 160 GeV.
The $\mu^+$ beam,  originated from the decay of $\pi$ and K mesons,
has a  longitudinal polarisation of about -76\%.
The target polarisation can be oriented both longitudinally and transversely
with respect to the beam direction.
Most of the time, data have been taken in
the longitudinal target spin mode, to measure 
$\Delta G/G$~\cite{Ageev:2005pq,brona}, the polarisation 
of the gluons in a longitudinally polarised nucleon; in parallel
very precise
$A_1^d$ data are also collected~\cite{Ageev:2005gh,santos,Alexakhin:2006vx}.
In about 20\% of the running time, data were taken with 
the transverse target polarisation 
to measure transverse spin effects.
The initial choice of the deuteron ($^6$LiD) as target material was due to the 
favourable value of the dilution factor ($f \simeq 0.4$),
particularly important for the measurement of 
$\Delta G/G$. 
Measurements with a NH$_3$ polarised proton target are 
also part of the experimental programme.

The uniqueness of the COMPASS experiment consists in its capability to
measure lepton-nucleon scattering with a high energy and high
intensity beam over a 
broad kinematical range, and with a large angular acceptance for the
final state particles.
To match these goals, the spectrometer~\cite{spectro} comprises 
two magnetic stages, 
which include state-of-the-art small and large area
tracking detectors, a RICH detector~\cite{rich}, hadronic
calorimeters, and systems for muon identification.

The first magnetic stage (the ``large angle spectrometer'') has a
design acceptance of about $\pm 200$ mrad in both planes,
to fully contain the hadrons of the current jet.
The second stage (the ``small angle spectrometer'') 
measures the higher energy forward particles.

The polarised target system~\cite{spectro,finger} consists of two oppositely 
polarised target cells,
60~cm long each and 3 cm diameter sitting in a superconducting
polarised target magnet (PTM), so that data are collected 
simultaneously for the two target spin orientations.
In the first three years of data taking, the large acceptance COMPASS
PTM was not available, and the PTM from the previous experiment SMC was
used, with a reduction in the  acceptance of the apparatus.
The PTM provides both a solenoidal 
field and a dipole field used for adiabatic spin rotation and for the transversity 
measurements.
To run in the transverse polarisation mode, the target 
is first polarised in the solenoidal field (typical values of 
the polarisation are $P_T \simeq 0.50$),
then the spins are frozen and rotated 
adiabatically to the transverse direction.

Since the spectrometer acceptance is somewhat different for the two target 
halves, the asymmetries cannot be obtained from a direct
comparison of the number of events collected in the two cells.
Usually, in a data taking period in the transverse mode,
the spins of both the target cells are reversed 
by exchanging the microwave frequencies of the cells after 
about 5-6 days of data taking.
To reduce possible systematic effects the  asymmetries have been
evaluated using
combinations of the number of events collected in the two cells
with opposite polarisation,  before and after the spin reversal.
In order to minimise possible efficiency variations 
before and after the spin reversal, during the data taking
in the transverse spin mode, particular care was put in keeping
the spectrometer functioning smoothly.

In the first three years of data taking from 2002 to 2004, COMPASS
collected about 200 TByte of raw data in the transverse configuration.
Thanks to the continuous improvement of the hardware and of the
reconstruction software, the number of reconstructed events increased
by about a factor of 2 each year.

 \section{COMPASS results}
The results for the Collins and Sivers asymmetries from the 2002 data 
have been published at the beginning of 2005~\cite{Alexakhin:2005iw}. 
In the mean time the 2003 and 2004 data have  been fully analysed, and the
paper has been sent for publication~\cite{newpap}
soon after this Conference.

Preliminary results for 
the Collins and Sivers asymmetries for identified particles
from the 2003 and 2004 data,
for the two-hadron asymmetry from the whole data set,
and for the $\Lambda$ polarisation from the 
2002 and 2003 data have been produced, and are also described in the following.

 \subsection{The Collins and Sivers asymmetries for charged hadrons}
 \subsubsection{Data analysis}

The 2003 and 2004 data have been analysed essentially in the same way
as the 2002 data.
Only DIS events (defined by 
a photon virtuality $Q^2 > 1$ (GeV/$c$)$^2$, and a
fractional energy of the virtual photon $0.1 < y < 0.9$)
with at least one reconstructed charged hadron produced
in the muon interaction point have been used.
The asymmetries have been evaluated for two (correlated) data
samples:\\
- the ``leading hadron'' sample, in which only the leading
(i.e. the most energetic) reconstructed hadron in the DIS event is used.
In our selection the leading hadron must have $z>0.25$.
The total statistics from the data collected
in the years 2002, 2003,
and 2004 is 1.4, 3.0, and 5.8 million  events.\\
- the ``all hadron'' sample, in which  all the
reconstructed hadrons with $z>0.2$ are used.\\
The reason for this ``double'' evaluation is that, in the string 
model~\cite{Artru}, it is expected that in the fragmentation
the sub-leading meson prefers the opposite side with respect to the
leading one, thus diluting the signal.

As already mentioned, the  Collins and Sivers asymmetries have been extracted
from the same data sample looking separately
at the modulation of the distribution
in the Collins and Sivers angles
$\Phi _C$ and $\Phi _S$.
Taking advantage of the two-cell configuration of the
polarised target system, we used as estimators the quantities
$$
A_j(\Phi_{j}) = \frac{ N_{j,u}^{+}(\Phi_{j})}
                     {N_{j,u}^{-}(\Phi_{j})} \cdot
\frac{ N_{j,d}^{+}(\Phi_{j})} 
                     {N_{j,d}^{-}(\Phi_{j})} 
 \simeq 1 + 4 \cdot A_j^m , \; \; \; \;  j=C,S \, .
$$
These estimators, used  as cross-check in the 2002 analysis, 
have several advantages.
In particular they minimise the possible acceptance effects,
and spin independent term in the cross-section cancel out at first order.

Because of the large statistics used to extract  the final asymmetries
(about a factor of 8 with respect to that of the 2002 data),
a large amount of work has been done to evaluate the systematic errors.
The data have been scrutinised, and the
stability of the spectrometer during the different
periods of data taking has been checked. 
Thanks to the fact that
the transverse spin measurements usually were  
carried out at the end of the data taking, when  the spectrometer was 
fully operational and functioning smoothly, only a small fraction
of the data had to be rejected.
Many tests have been performed on the stability of the results
vs time and acceptances, and on the consistency of the results
from the different data taking periods.
In all the tests no indication
for systematic effects has been observed, with the
conclusion that  the systematic 
errors are considerably smaller than the statistical ones.

\subsubsection{Results}

\begin{figure}[htb] %
\begin{center}
\includegraphics[width=0.8\textwidth]{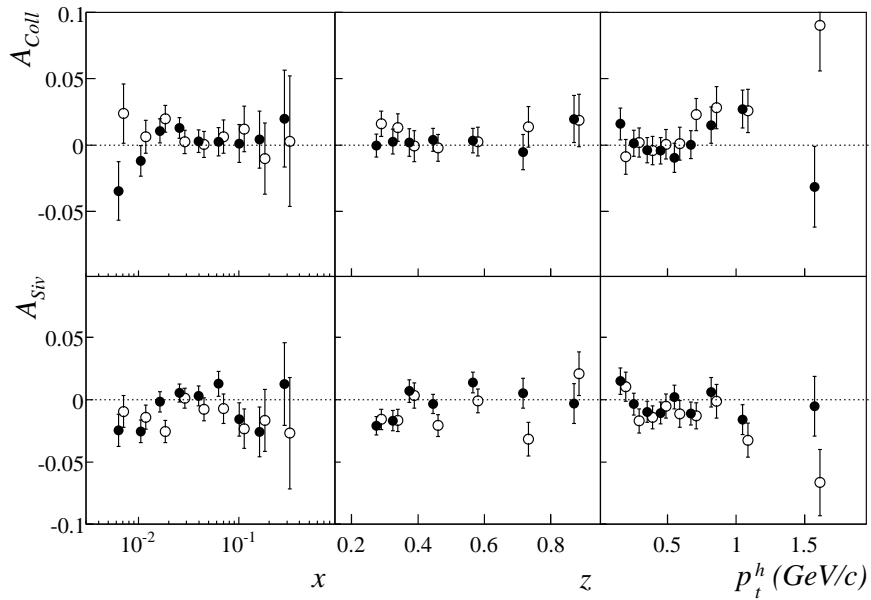}
\end{center}
\vspace*{-0.8cm}
\caption{Collins asymmetry (top) and Sivers asymmetry 
(bottom) against the Bjorken variable $x$ (or $x_{Bj}$), $z$ and 
$p_T^h$ for positive (full circles) and negative 
leading hadrons (open circles) from the 2002, 2003, and 2004 data.}
\label{fig:r0234l}
\end{figure}

\begin{figure}[hbt] %
\begin{center}
\includegraphics[width=0.8\textwidth]{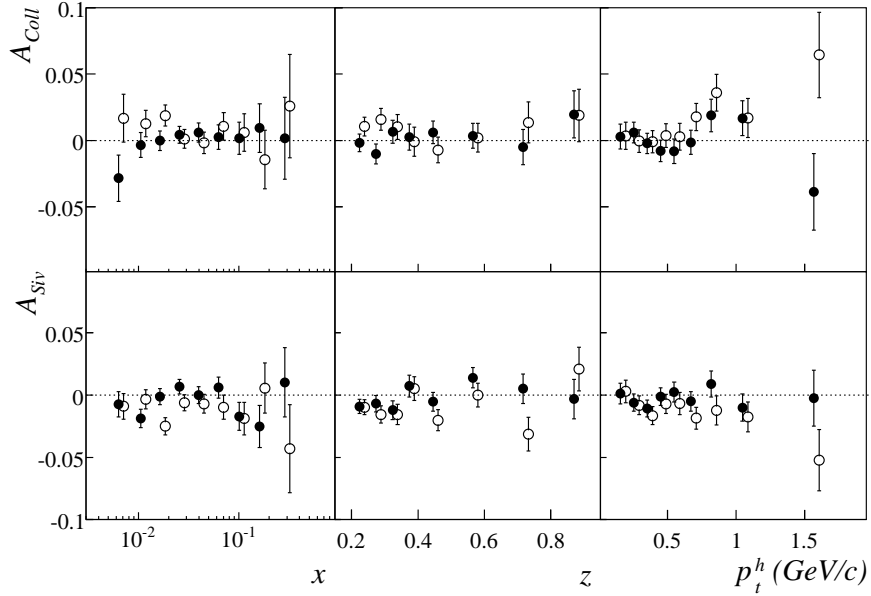}
\end{center}
\vspace*{-0.8cm}
\caption{Collins asymmetry (top) and Sivers asymmetry 
(bottom) against $x$, $z$ and 
$p_T^h$ for all positive (full circles) and all negative 
hadrons (open circles) from 2002 - 2004 data.}
\label{fig:r0234a} 
\end{figure}

The  results from the 2002-2004 data~\cite{newpap}
for the leading hadron sample and for the all hadron sample are given in
Fig.~\ref{fig:r0234l} and~\ref{fig:r0234a} respectively.
They confirm with much better precision
the results from the 2002 data: all the asymmetries are 
compatible with zero, within the statistical errors, now of the
order of  percent.

The results for the Collins asymmetry are at variance 
with the HERMES~\cite{Diefenthaler:2005gx} measurement on the  
transversely polarised proton target.
This is true not only at small values of the Bjorken variable $x$,
where anyhow the transversity distributions are expected to be small,
but also in the range  $0.05<x<0.3$, where
HERMES observed  asymmetries  different from zero,
both for positive and negative pions.
Moreover, the fact that our results are obtained for unidentified hadrons
cannot justify the difference since most of charged hadrons 
are in fact pions
(see also next section).
Thus, the obvious conclusions is that the asymmetries compatible with
zero measured
by COMPASS are due to the use of a deuteron target. i.e. to
the different combination of the transversity PDF's
we are probing.
In a simplified parton model framework, assuming that only pions 
are involved and neglecting the contribution of the sea quarks,
in the case of a deuteron target it is
$$
A_{Coll}^{d, \pi^+}  \simeq  
\frac{\Delta_{T} u_v  + \Delta_{T} d_v}{u_v + d_v }
\frac{4 \Delta_{T}^0 D_1 + \Delta_{T}^0 D_2}
     {4 D_1 + D_2} \, , \; 
A_{Coll}^{d, \pi^-}  \simeq 
\frac{\Delta_{T} u_v  + \Delta_{T} d_v}{u_v + d_v }
\frac{\Delta_{T}^0 D_1 + 4 \Delta_{T}^0 D_2}{D_1 + 4 D_2} 
$$
where $ D_1 = D_u^{\pi^+} = D_d^{\pi^-}$ and 
$ D_2 = D_d^{\pi^+} = D_u^{\pi^-}$ are the favoured and unfavoured FF's.
For a proton target, the asymmetries are given by
$$
A_{Coll}^{p, \pi^+}  \simeq 
\frac{4 \Delta_{T} u_v \Delta_{T}^0 D_1 + \Delta_{T} d_v \Delta_{T}^0 D_2}
     {4  u_v D_1 + d_v D_2} \, , \; 
A_{Coll}^{p, \pi^-}  \simeq 
\frac{4 \Delta_{T} u_v \Delta_{T}^0 D_2 + \Delta_{T} d_v \Delta_{T}^0 D_1}
     {4  u_v D_2 + d_v D_1} \, .
$$
Neglecting the $d$ quark contribution to the asymmetries, the HERMES data
suggest $\Delta_{T}^0 D_2 \simeq - \Delta_{T}^0 D_1 $, a somehow unexpected
result which is not excluded from the BELLE 
measurements~\cite{Abe:2005zx,Ogawa:2006bm}.
Even in this case, however, it is not straightforward to understand why 
the deuteron asymmetries are so small.

\begin{figure}[hbt] %
\begin{center}
\includegraphics[width=.6\textwidth]{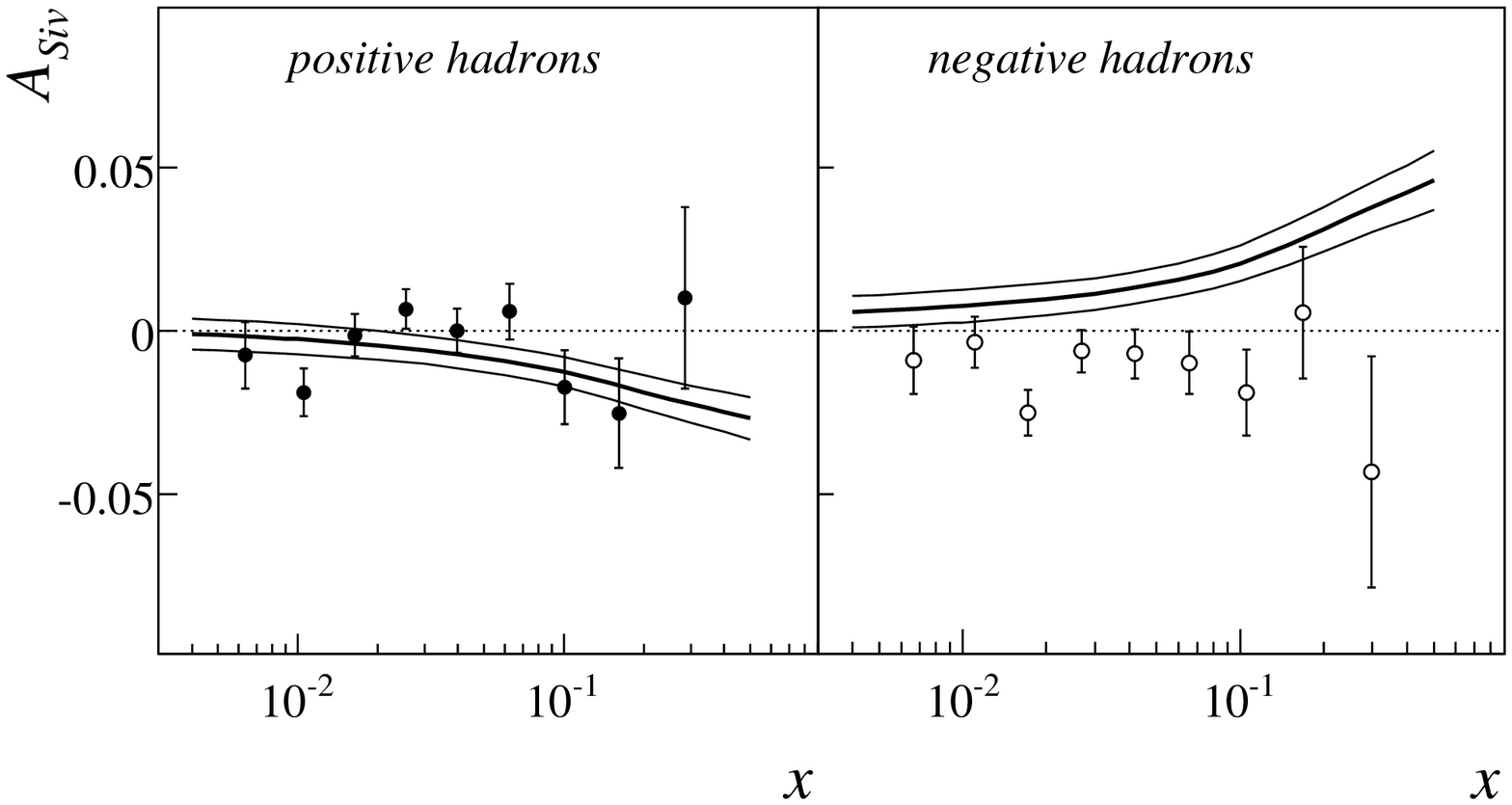}
\includegraphics[width=.6\textwidth]{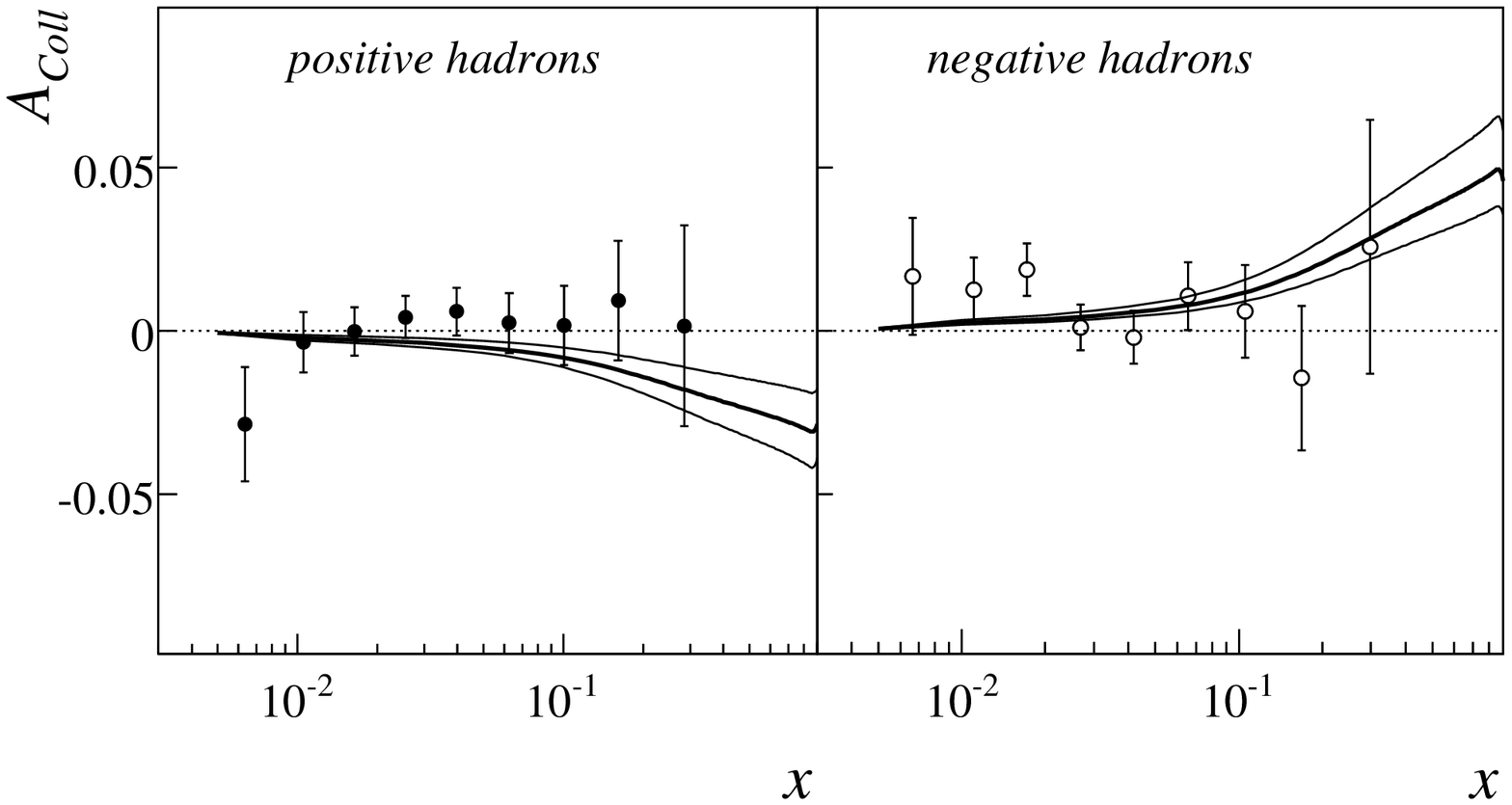}
\end{center}
\vspace*{-0.8cm}
\caption{Collins asymmetry: comparison between the 2002-2004 COMPASS results for positive 
(left) and negative (right)
hadrons with the calculations of Ref.~\cite{vy} (upper plots) and Ref.~\cite{efre} (lower plots).
The upper and lower curves represent the theoretical uncertainty.
}
\label{fig:collefr} 
\end{figure}

Phenomenological calculations have been performed by three different 
groups~\cite{vy,efre,anse} 
to extract information on the transversity PDF's and the Collins 
FF's.
The common approach has been to first fit the HERMES 
data~\cite{Airapetian:2004tw,Diefenthaler:2005gx}
 and then compare the model calculations with the 
BELLE
 and COMPASS 2002 measurements~\cite{Alexakhin:2005iw}.
All the considered data are in agreement with 
the result that the favoured and unfavoured Collins FF's
are of the same magnitude and of opposite sign, and that the $u$ quark 
contribution is dominant. 
The comparison between our new data and the calculations of Ref.~\cite{vy} (upper plots)
and Ref.~\cite{efre} (lower plots)
is shown in Fig.~\ref{fig:collefr}.
\begin{figure}[hbt] %
\begin{center}
\includegraphics[width=.6\textwidth]{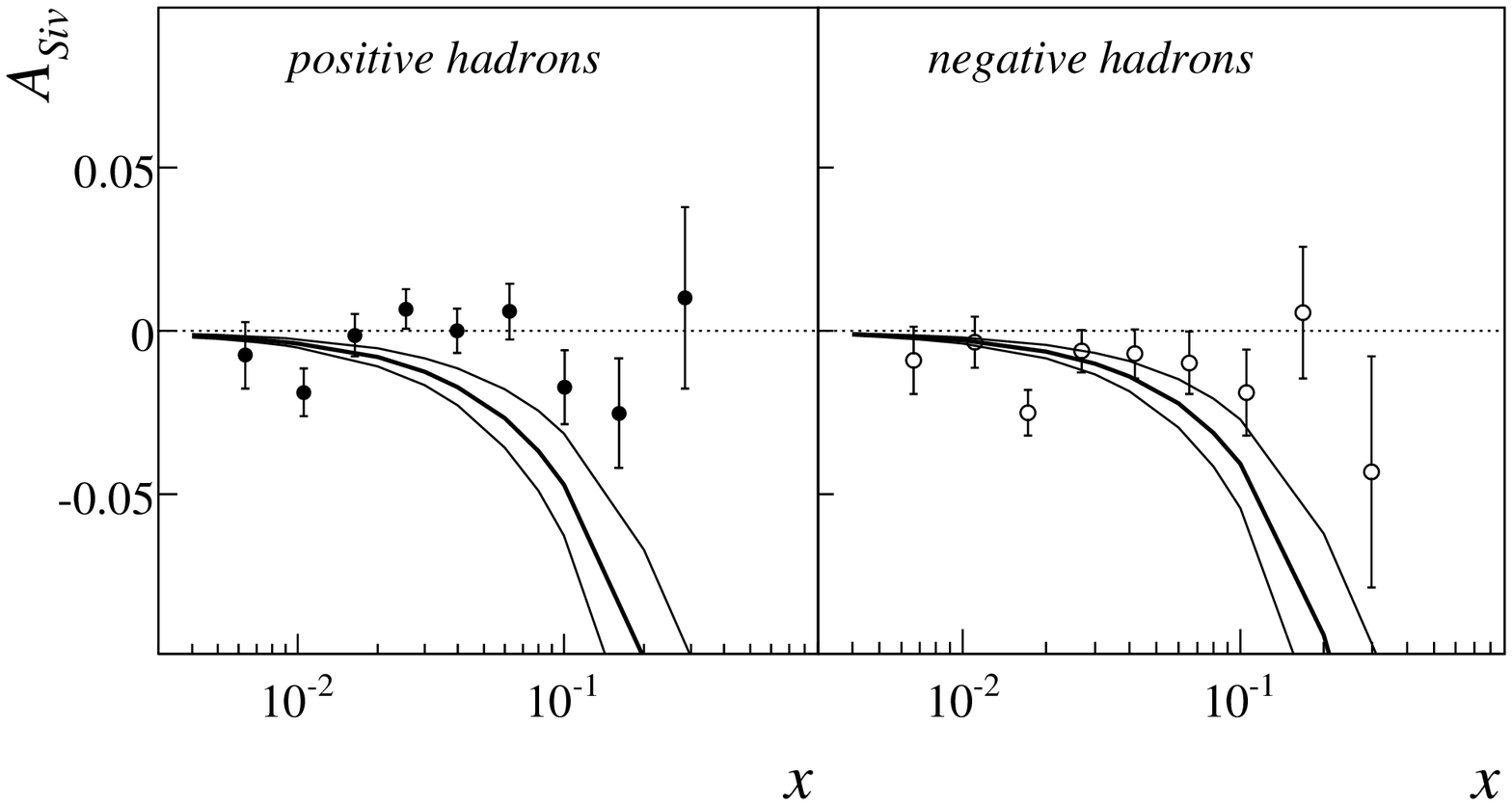}
\includegraphics[width=.6\textwidth]{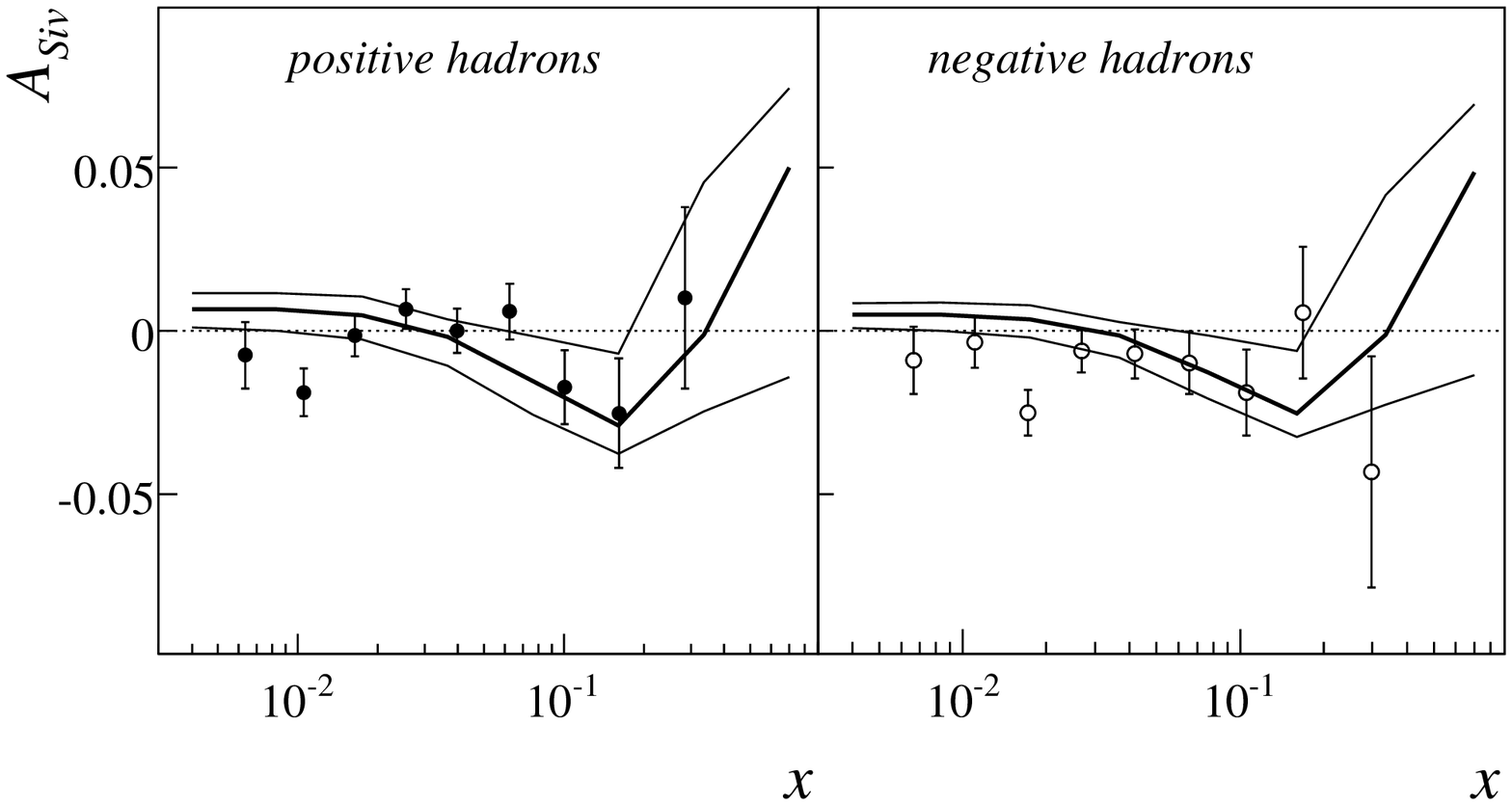}
\end{center}
\vspace*{-0.8cm}
\caption{Sivers asymmetry: comparison  between the 2002-2004 COMPASS results for positive 
(left) and negative (right)
hadrons with the calculations of Ref.~\cite{vy} (upper plots) and Ref.~\cite{sanse} (lower plots).
}
\label{fig:sivans} 
\end{figure}
The agreement is still satisfactory, but clearly
a global analysis including
the present deuteron data will allow 
to put tighter constrains on the transversity distributions.
It is also clear, however, that more data from BELLE, HERMES, and
COMPASS (in particular with a proton target), are needed to 
extract precise information. 

The new precise  results for the Sivers asymmetry  
confirm that also this effect on a deuteron target at COMPASS
energies is very small, while the measurements on a proton
target from HERMES~\cite{Airapetian:2004tw,Diefenthaler:2005gx}
give a clear indication for a positive asymmetry for $\pi^+$
and values compatible with zero for $\pi^-$.
In the simplified parton model described before,
the Sivers asymmetries are given by
$$
A_{Siv}^{p, \pi^+}  \simeq 
\frac{4 \Delta_0^{T} u_v D_1 + \Delta_0^{T} d_v D_2}
     {4  u_v D_1 + d_v D_2} \, , \;
A_{Siv}^{p, \pi^-}  \simeq 
\frac{4 \Delta_0^{T} u_v D_2 + \Delta_0^{T} d_v D_1}
     {4  u_v D_2 + d_v D_1} \, .
$$
and 
$$
A_{Siv}^{d, \pi^+}  \simeq   A_{Siv}^{d, \pi^-}  \simeq 
\frac{\Delta_0^{T} u_v  + \Delta_0^{T} d_v}{u_v + d_v}  \, .
$$
The COMPASS result thus indicates a cancellation between the $u$ 
and $d$ quark Sivers functions while the HERMES results
suggest $\Delta_0^{T} d_v  \simeq -2\Delta_0^{T} u_v $.

Much more complete theoretical calculations have been 
done~\cite{vy,sanse,Collins:2005ie,Anselmino:2005an} to constrain the
Sivers functions, using mainly the HERMES data.
In all cases, the result is that the $u$ and $d$ quark PDF
are of the same size and have opposite sign, and
their parametrisations give asymmetries for the deuteron which
are in agreement COMPASS 2002 results.
The agreement with the new COMPASS data is not always as good as for the
2002 results, 
as can be seen in Fig.~\ref{fig:sivans},
in which the curves are the calculations from Ref.~\cite{vy} (upper plots)
and Ref.~\cite{sanse} (lower plots).
It is clear that our results will allow to better
define the overall picture.

 \subsection{The Collins and Sivers asymmetries for identified hadrons}
The RICH detector was fully operational during transverse spin data taking
in the years 2003 and 2004.
For these data, the full analysis including particle 
identification  has been performed to measure $\pi$ and $K$ asymmetries.
Momentum cuts have been applied to be above the Cherenkov threshold
($p_{\pi} \simeq 2$ \gevc\ and $p_{K} \simeq 9$ \gevc ) and to
guarantee at least a 1.5 $\sigma$ mass separation between the two
hypothesis ($p_{\pi , K} <$ 50 \gevc ). 
The statistics of the final all hadron samples is 
5.2 and 4.5 millions of $\pi^+$ and $\pi^-$, and
0.9 and 0.45 millions of $K^+$ and $K^-$.

\begin{figure}[bt] %
\begin{center}
\includegraphics[width=0.8\textwidth]{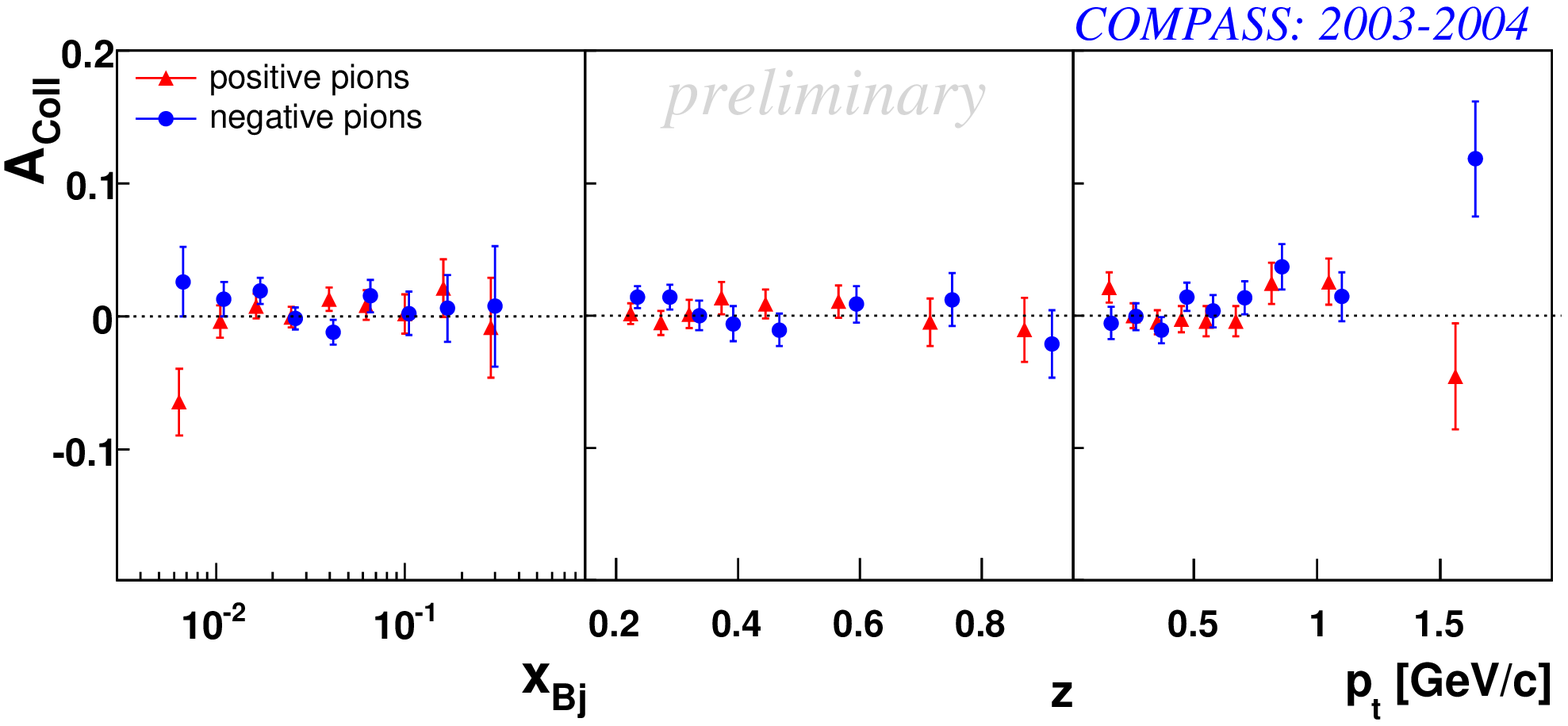}
\includegraphics[width=0.8\textwidth]{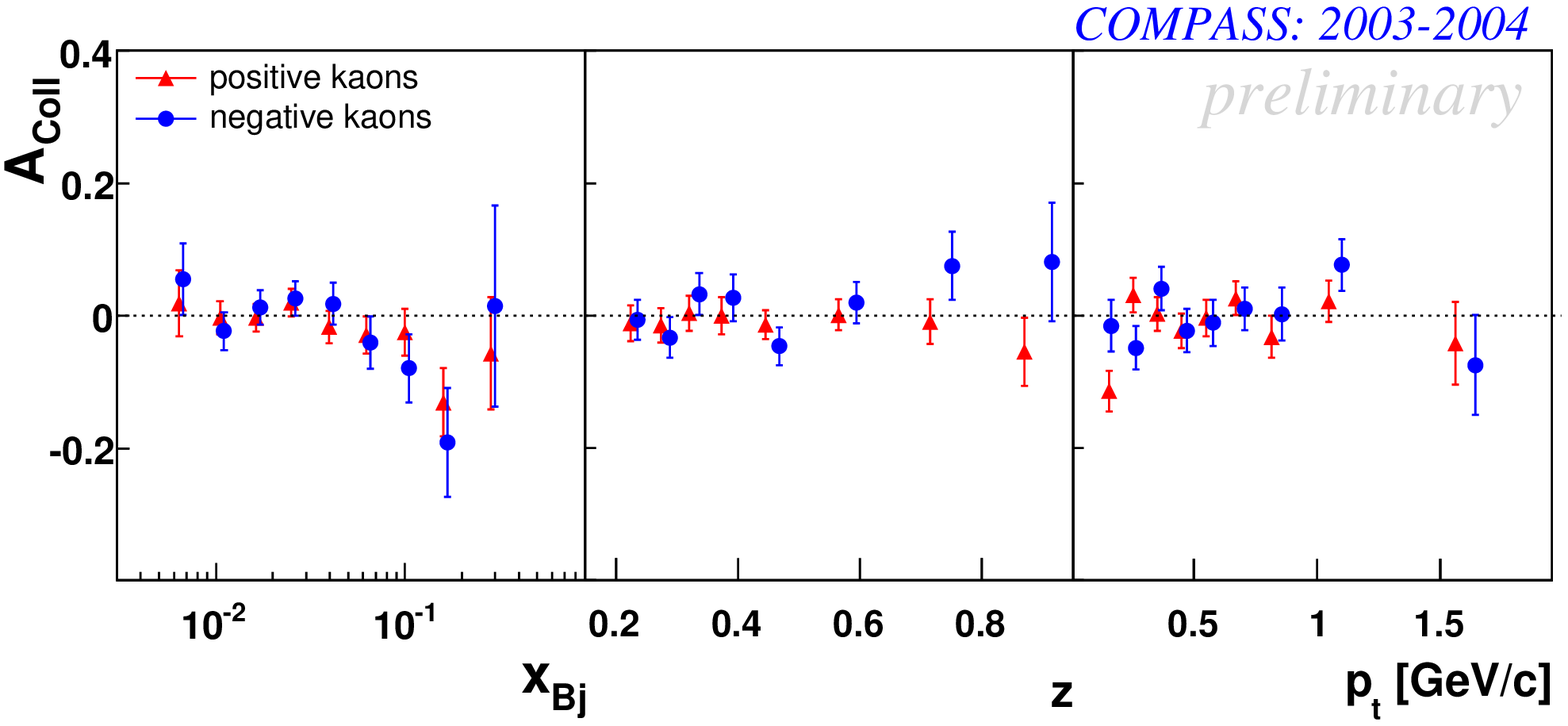}
\end{center}
\vspace*{-0.8cm}
\caption{Collins asymmetry against $x$, $z$ and 
$p_T^h$ for positive (full triangles) and negative  (full circles)
 $\pi$ in the upper plots, and $K$ in the lower plots, from 2003 and 2004 data.}
\label{fig:cpkar034}
\end{figure}

\begin{figure}[bt] %
\begin{center}
\includegraphics[width=0.8\textwidth]{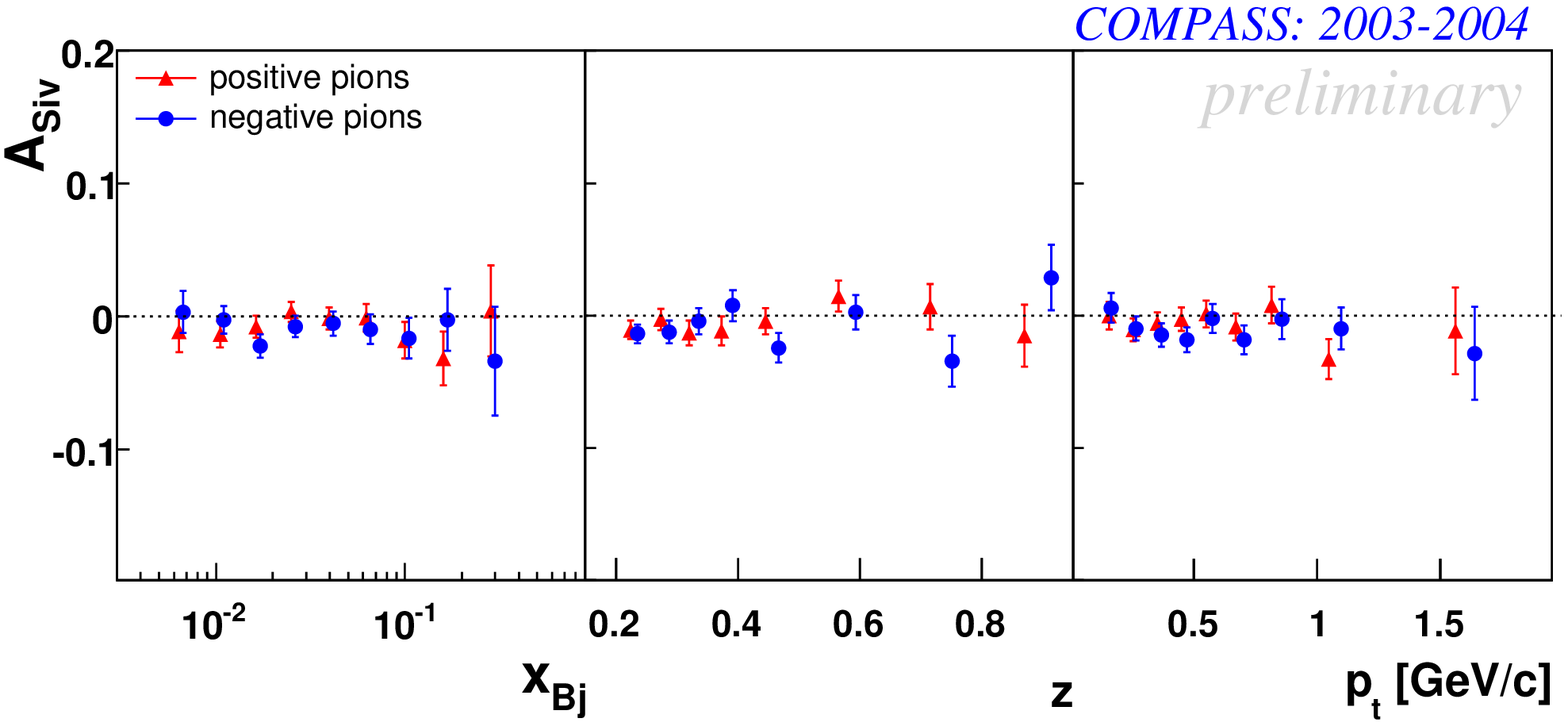}
\includegraphics[width=0.8\textwidth]{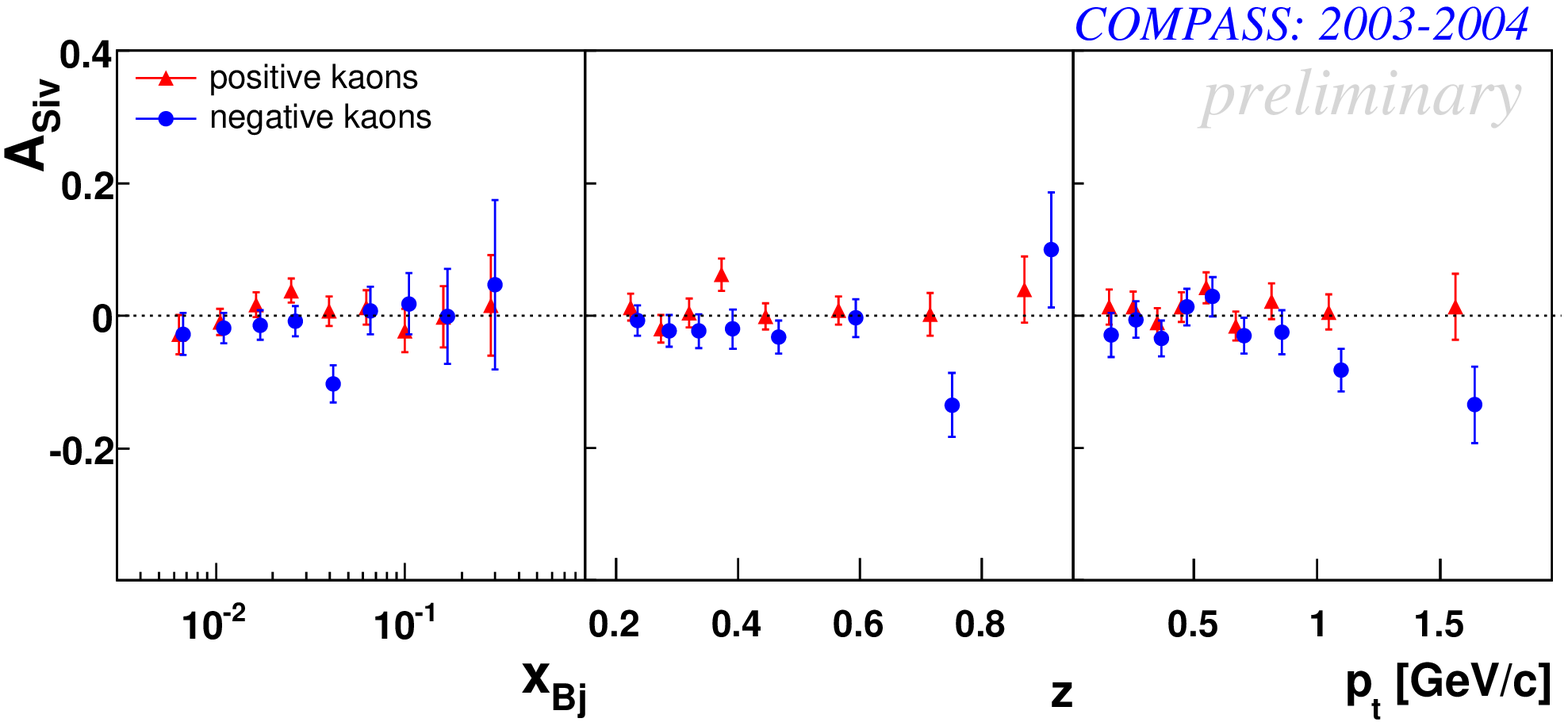}
\end{center}
\vspace*{-0.8cm}
\caption{Sivers asymmetry against $x$, $z$ and 
$p_T^h$ for positive (full triangles) and negative  (full circles)
 $\pi$ in the upper plots, and $K$ in the lower plots, from 2003 and 2004 data.}
\label{fig:spkar034}
\end{figure}

The Collins and Sivers asymmetries have been evaluated both for the
all hadron and the leading hadron samples, getting in the two cases
very similar results.
The Collins asymmetries for all pions and all kaons
are shown in Fig.~\ref{fig:cpkar034} against $x$, $z$ and 
$p_T^h$.
The corresponding Sivers asymmetries
are given in Fig.~\ref{fig:spkar034}.
The errors shown in the figures are statistical only
(all the tests performed following the charged hadron analysis
have shown no evidence for systematic effects) and they are 
comparable with the HERMES statistical errors in the overlapping 
$x$ range, i.e. in the last five bins.

As expected, the measured pion asymmetries are very similar to the 
unidentified hadron asymmetries,
thus confirming the picture described in the previous section.

The kaon asymmetries have somewhat larger errors, and again no significant
signal is visible both in the Collins and the Sivers asymmetries.
The recent results on proton from HERMES~\cite{hdis06}
show a clear positive asymmetry only
in the Sivers case, for positive $K$, and the
COMPASS result could again indicate a cancellation due to the
deuteron target.
Both the HERMES and
COMPASS data are new, and the interpretation work is ongoing.

 \subsection{Two-hadron asymmetry}

Preliminary results from the 2002 and 2003 data have  been 
produced last year. 
Now the analysis is almost over and the results from the 2004 data,
which double the available statistics, are presented.

The analysis has been performed much as the for the single hadron asymmetries 
measurements, the main difference being that now a hadron pair in the 
final state has to be identified.

\begin{figure}[bt] %
\begin{center}
\includegraphics[width=1\textwidth]{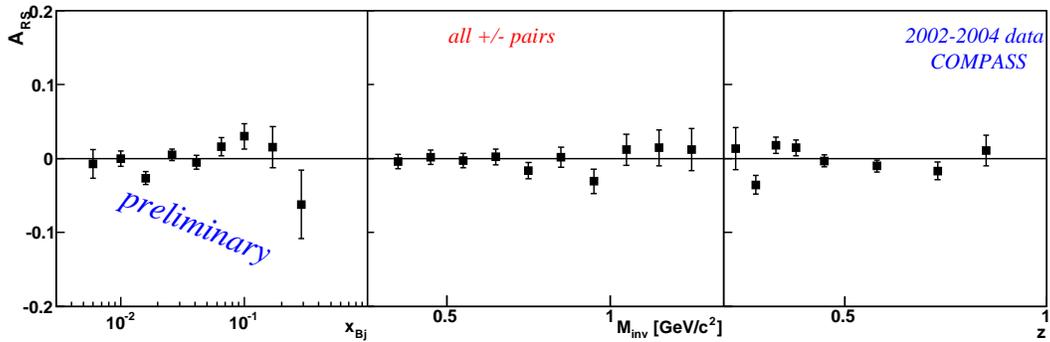}
\end{center}
\vspace*{-0.7cm}
\caption{Two hadron asymmetry against $x$, the invariant mass 
of the hadron pair $M_{inv}$, and 
$z$ for all oppositely charged hadron pairs
from 2002, 2003 and 2004 data.}
\label{fig:2ha0234}
\end{figure}

The ``standard'' recepi is to consider all combinations of
positive - negative hadrons, labelling as ``hadron 1'' the positive one
in the angle $\phi_{RS}$ calculation.
Only pairs with $z_1 > 0.1 , \, z_2 > 0.1$ and $z =z_1 + z_2 < 0.9$ have been 
used.
The total number of pairs for the whole data sample consists
of about 6 million pairs.
The preliminary results for $A_{RS}$ against $x$, against the invariant mass 
of the pair $M_{inv}$, and against $z$ are shown in Fig.~\ref{fig:2ha0234}.
The error bars show the statistical error. 
Once more, the asymmetries are compatible with zero,
within the small statistical error.
This is at variance with the HERMES measurement for a proton 
target~\cite{2hhermes} of
a significantly different from zero two-hadron asymmetry.
The COMPASS result may be a further evidence of the cancellation 
which takes place in the case of the deuteron target.
Theoretical work on the interpretation and the modelling of the
two hadron asymmetry 
is ongoing~\cite{Bacchetta:2006un}, and, in spite of the still unknown FF's
which appear in its expression, it looks to be a very promising tool to
better understand transversity.

\begin{figure}[tb] %
\begin{center}
\includegraphics[width=0.8\textwidth]{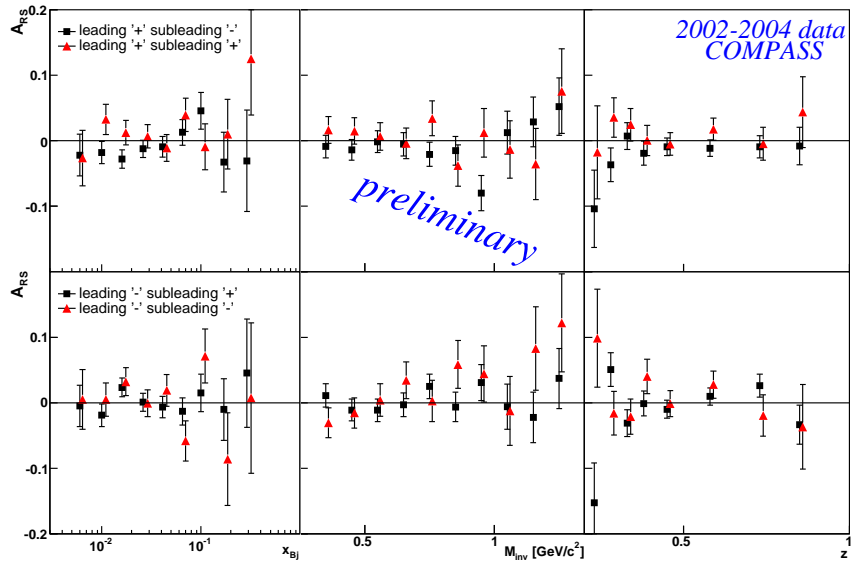}
\end{center}
\vspace*{-0.7cm}
\caption{Two hadron asymmetry against $x$, the invariant mass 
of the hadron pair $M_{inv}$ and 
$z$ for ``$z$-ordered'' hadron pairs
from 2002, 2003 and 2004 data.}
\label{fig:2hz0234}
\end{figure}
Several attempts to isolate a signal in the COMPASS data
have been done looking at different definitions of the hadron pair.
In particular in each event we considered only the two hadrons with the
largest $p_T$,
to look at hard DIS processes, measuring again asymmetries
compatible with zero~\cite{Dubna05}.
An other approach consists in selecting
only the leading and subleading 
produced particles (``$z$-ordered'' pairs) in the event, which should be 
the most favourable combination following the string model.
The preliminary results are shown in Fig.~\ref{fig:2hz0234};
also in this case it is hard to see a signal.
Work to measure the asymmetries using the RICH information is in progress,
and preliminary results should be available soon.

 \subsection{$\Lambda$ polarimetry}
The analysis of the 2004 data is still ongoing, and the preliminary
results on $\Lambda$ polarisation
discussed here are from the 2002 and 2003 data.

\begin{figure}[bt] %
\hfill
\includegraphics[width=0.45\textwidth]{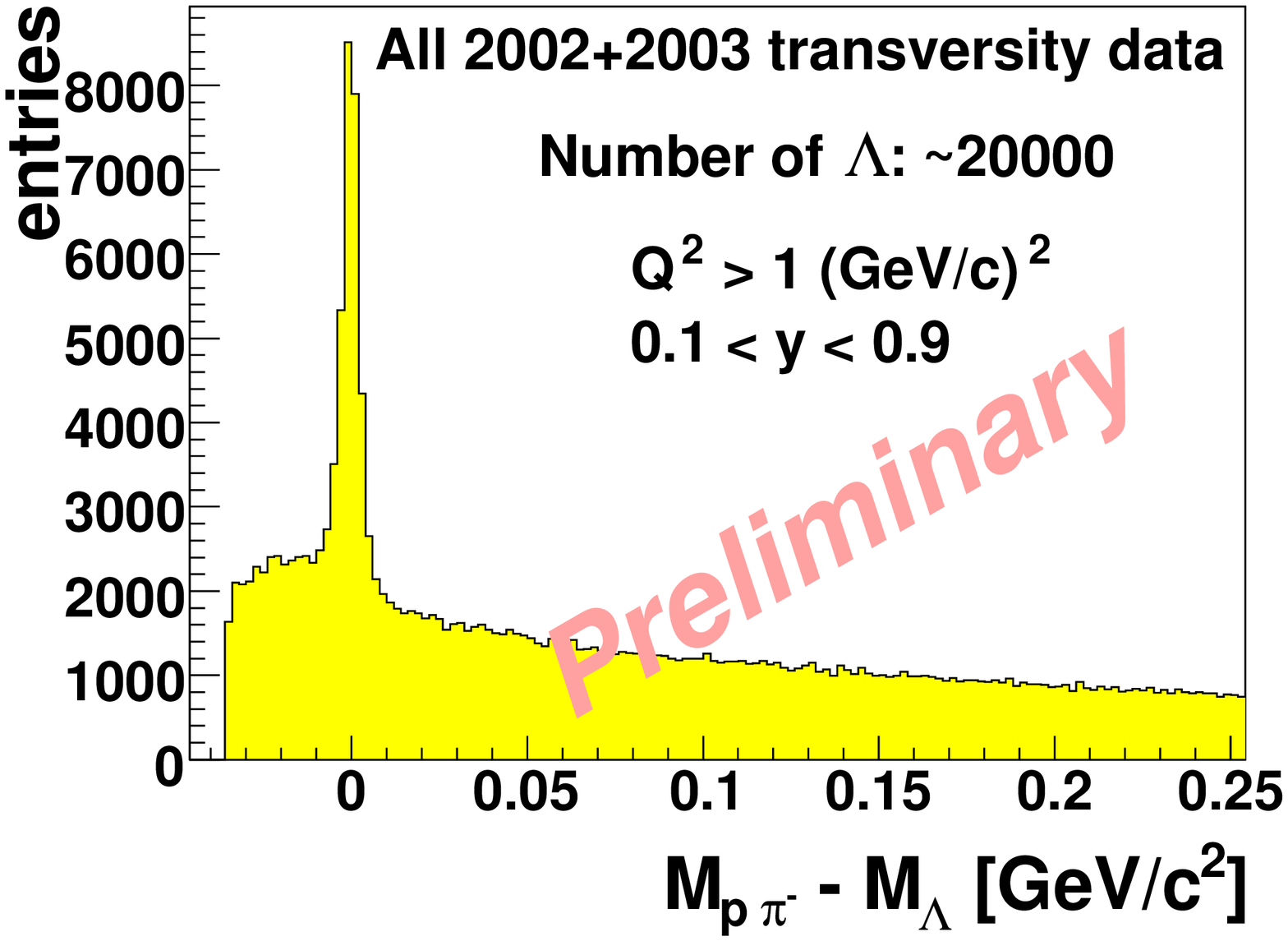}
\includegraphics[width=0.53\textwidth]{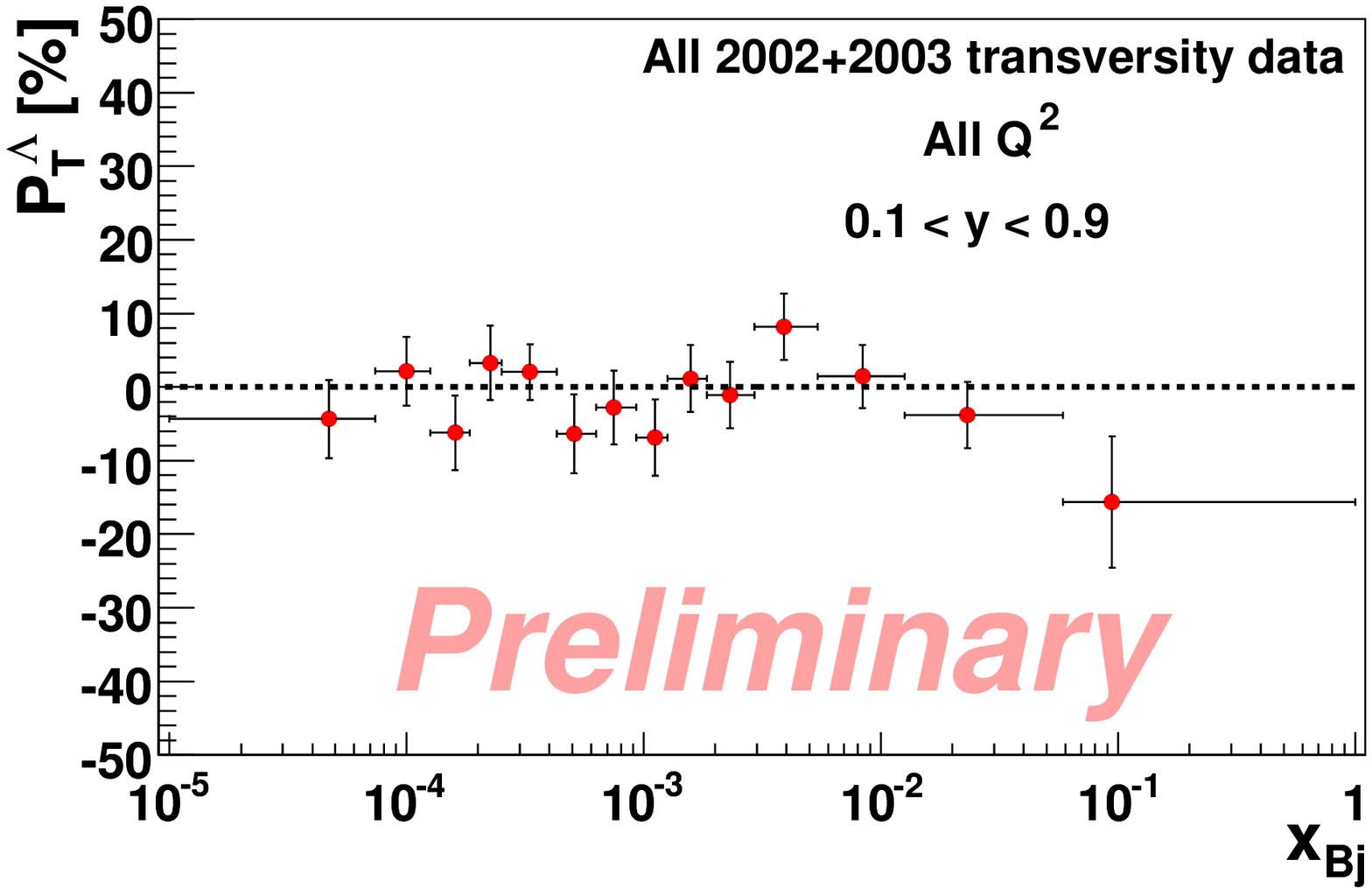}
\hfill
\vspace*{-0.3cm}
\caption{Left: invariant mass spectrum for all DIS events with 
$Q^2>1$ (\gevc )$^2$.
Right: transverse $\Lambda$  polarisation for all $Q^2$ events. 
 }
\label{fig:Lambda}
\end{figure}

Presently, the RICH information is not used in the $\Lambda$ identification,
which relies on kinematical cuts~\cite{aferrero}.
The background is mainly due to $K^0$ decays, photon conversion, and
wrongly reconstructed decay vertices.
To reduce these last two contributions the decay vertex is required to fall
in between the downstream-end of the target and the first tracking detector (about 1.1
m downstream of the target).
Also, cuts on the transverse momentum of the positively charged
secondary particle with respect to the $\Lambda$ candidate direction
are applied.
As can be seen in Fig.~\ref{fig:Lambda} (left plot) the invariant mass spectrum
shows a very clean signal over a smooth  background.
The number of $\Lambda$'s inclusively produced is about 20k 
for $Q^2>1$ (\gevc )$^2$ and 10 times more for all $Q^2$. Similar numbers 
are expected from the 2004 data alone.
The number of $\bar{\Lambda}$'s is about a factor of 2 smaller.

The preliminary result for the transverse $\Lambda$  polarisation 
from the 2002 and 2003 data is shown in Fig.~\ref{fig:Lambda} (right)
for all $Q^2$ events. 
Given the correlation between $x$ and $Q^2$,
the contribution of the events with $Q^2>1$ (\gevc )$^2$ is
dominant in the last two $x$ bins.
The  errors on the data points
are the statistical ones (systematic errors have been
evaluated to be not larger than the statistical errors);
as expected, they are much larger than for the hadron asymmetries.
More input on this interesting channel will come from the 2004 data
and from the $\bar{\Lambda}$ polarisation~\cite{aferrero2}.

 \section{Conclusions}
The data collected by the COMPASS experiment with the transversely
polarised deuteron target in the years 2002-2004 have been
almost completely analysed.

All the transverse spin effects investigated in so far
(Collins and Sivers asymmetries, two-hadron asymmetries,
${\Lambda}$ polarisation) are compatible with zero and can
be interpreted as an indication of a cancellation between the $u$ and the $d$
contributions in the deuteron.

Important theoretical work is ongoing to extract informations
on the transversity PDF's from the new measurements of the Collins
asymmetries by
COMPASS and HERMES, and of the Collins FF's by BELLE.

A first interpretation of the measured Sivers asymmetries in terms
of PDF's has also been put forward and the new  data will help
in better specifying the global picture.

The general feeling is that 
much more data and a global analysis of all available measurements
are needed to formulate a more quantitative picture.
In addition to the finalisation of the analysis of the 2002-2004 data
the COMPASS contribution in the near future will be the measurement,
with comparable statistics, of the transverse spin effects 
with the proton target, now scheduled at the beginning of the 2007 run.

On a longer time scale, plans
are being defined to continue this promising and exciting field of research
in SIDIS at CERN~\cite{stratg}.
 \bigskip



 \end{document}